# Using Multiple Beams to Distinguish Radio Frequency Interference from SETI Signals

G. R. Harp

Allen Telescope Array, SETI Institute, 515 N. Whisman Rd., Mountain View, CA 94043

**Abstract**

The Allen Telescope Array (ATA) is a multi-user instrument and will perform simultaneous radio astronomy and radio SETI (search for extra-terrestrial intelligence) observations. It is a multi-beam instrument, with 16 independently steerable dual-polarization beams at 4 different tunings. Here we describe a new method for identifying radio frequency interference (RFI) that leverages the unique attributes of the ATA. Given 4 beams at one tuning, it is possible to distinguish RFI from true ETI signals by pointing the beams in different directions. Any signal that appears in more than one beam can be identified as RFI and ignored. We discuss the effectiveness of this approach using realistic simulations of the fully populated 350 element configuration of the ATA as well as the interim 32 element configuration. Over a 5 minute integration period, we find RFI rejection ratios exceeding 50 dB over most of the sky.

**Introduction**

Radio frequency interference (RFI) is a growing problem for all radio astronomy applications, but is especially problematic in the search for extraterrestrial intelligence (SETI) (Ekers 2002, Davis 2004). A key element of any RFI mitigation strategy is to discriminate RFI from naturally occurring signals, or in the case SETI, from artificial signals originating outside our solar system. Once the RFI is identified, corrective action can be taken. One approach keeps an ongoing database of RFI signals as they are identified, and abandons frequency ranges where RFI is recently or persistently observed. This approach is currently implemented in the signal processing system employed for the SETI Institute's Project Phoenix (Ekers 2002, Backus 2004).

In this paper we examine a new method of RFI discrimination that is especially appropriate for SETI observations and based on correlating the signals arriving from multiple single-pixel beams of an interferometer telescope. Two or more beams are pointed in different directions on the sky. If a signal appears strongly in more than one beam it must be RFI since identical signals could never appear from two different stars. This is just a generalization of an approach customarily used in SETI observations with single dish telescopes. When a candidate signal is found at one sky direction, one points the dish away from the target to see if the signal disappears (on/off test). To complete the analogy with the present approach, search efficiency could be improved by choosing a second target star for the "off" position. Because here we work with an interferometer, the on and off measurements are made simultaneously.

To quantify the effectiveness of multibeam RFI identification, we must examine the beam sidelobe pattern through which the RFI enters the telescope. As a concrete example we consider the sidelobe pattern synthetic beams at the Allen Telescope Array (ATA). The Allen Telescope Array (ATA) is a radio interferometer under construction at the Hat Creek Radio Observatory in Northern California. Each ATA element is a 20' diameter offset Gregorian telescope, and is operable over 0.5-11.2 GHz. The ATA will be constructed in three stages comprising 32, 206, and 350 elements at each stage. We simulate the synthetic beam patterns of the 350 element ATA (ATA-350) and the 32 element ATA (ATA-32) and examine the sidelobes through which RFI may enter the synthetic beam. From the statistical distribution of sidelobe levels, we estimate the probability that an RFI signal will enter strongly into one beam while being weaker or absent in all others. This is the probability of a "false positive," or that a bit of RFI masquerades as an ETI signal. We find that multi-beam discrimination is an effective way to identify RFI. Once identified, RFI can be eliminated from subsequent follow-up protocols thereby increasing search speed.

**Description of the Calculations**

We simulate observations where multiple synthetic beams are formed within the primary beam of a single antenna. The RFI is assumed to enter in a sidelobe of the primary beam because we avoid pointing at known RFI and because the primary beam (FWHM of 3.5-0.35° between 1-10 GHz) represents only a small fraction of the sky. Although the primary sidelobe pattern varies from high to low on a scale of half the primary beam width, we shall assume that this multiplicative factor does not change the statistical behavior of the synthetic beam sidelobe level (i.e. the sidelobe statistics are the same as for an isotropic antenna). This seems reasonable since all synthetic beams are within one beam width of one another.

Figure 1 displays the antenna layout for ATA-350 (left) and a beam pattern (right). The white dot at the center of the pattern is the synthetic beam peak, and the white circle indicates the half-power point of the antenna primary beam. Outside a few synthetic-beam widths of the maximum, the statistical distribution of sidelobes does not vary with position in the beam pattern. More specifically, if you look at the sidelobe distribution in a small area of the beam pattern centered around a direction $\hat{k}$, where $\hat{k}$ is far from the pattern maximum, then this local probability distribution is independent of $\hat{k}$, apart from the expected statistical fluctuations inherent in measuring probabilities over a finite area of the beam pattern. We find this to be quantitatively true in all our calculations.

In the calculations that follow, beam patterns are calculated on a square grid with ~4 million points over an angular range that does not include the beam maximum. A histogram of sidelobe power is accumulated, which when normalized to the number of grid points, gives an estimate of the probability density $P(s)$ of finding a sidelobe with level $s$. Such a histogram is displayed in fig. 2 which shows the levels in a "snapshot" observation.

Using these data we calculate the probability that RFI will appear as a "false positive" ETI signal by using the following trick. We place the synthetic beam maximum on the RFI and the observation beams in the far out sidelobe region. This is justified by the inversion symmetry of the beam pattern: for a beam placed on a source, the sidelobe power for the RFI is the same as the sidelobe power on the source when the beam is placed on the RFI. We then compare the sidelobe levels at the positions of the different beams and set a rejection threshold of $N$ dB for a false positive event. If one beam has a sidelobe level $N$ dB higher than all the others, this is a false positive. For $M$ observation beams, the probability of false positive $P_M$ (a.k.a. rejection ratio) is calculated from:

$$P_M(n) = M \int_0^1 ds\, P(s) \left\{ \int_0^{s/n} ds'\, P(s') \right\}^{M-1}, \quad (1)$$

where $n = 10^{-N/10}$. The term in curly brackets is the probability that a given beam will have a level less than or equal to $(s/n)$. The prefactor $M$ appears because we don't care in which beam displays the strong RFI. This equation is derived in the appendix.

Figure 3 shows a plot of $P_4(N)$ calculated from the data in figure 2. In most of our calculations we use $M = 4$ because this is the natural number of independent beams produced at the ATA for a given frequency tuning.

There is one more subtlety to consider. In the version of the SETI search system belonging to the SETI Institute, a single point on the sky is observed for ~5 minutes before moving on to the next point or next frequency. During this time, the observed signal is Fourier transformed to obtain the frequency power spectrum, which is then examined for characteristic ETI signals. It is not possible to perform a direct Fourier transform (FT) of all 5 minutes of data. Instead, one-second windows are FT'd and the resulting windows are integrated incoherently over the observation period (Backus 2004). This feature of the analysis greatly improves the rejection ratio since, as the source moves across the sky, the RFI (assumed fixed) moves through the beam sidelobes. Averaging over many sidelobes both narrows and heightens the distribution of $P(s)$. The degree of averaging depends on the source position and RFI position.

For simplicity, we put the RFI on the ground due east of the array. The antenna primary beam is placed at various declinations and the array is assumed to be at latitude 41° (where the ATA is located). The probability distribution is calculated by generating 300 beam patterns simulating 1 second integrations over a 5 minute observation period. For each source position, the sidelobe level power is averaged over all patterns. After averaging, the probability density is calculated as before.

Figure 4 shows $P(s)$ for a 5 minute track at declination 20° near transit. The sidelobe distribution is substantially narrowed as compared with fig. 2. The peak of the distribution is also about 10 times higher, but this rise is offset by the choice of a smaller bin size in fig. 4 as compared with fig. 2.

Such averaging leads to a greatly improved rejection ratio as shown by the blue curve in fig. 5 (c.f. fig. 3). We find that the chances are less than 1 in $10^5$ that the RFI will appear only 3 dB higher in one beam than in all the others.

**Results**

We begin by examining the sensitivity of the rejection ratio to the number of beams used. Figure 5 plots the rejection ratio for 2, 3 and 4 beams as determined from the data in fig. 4. Using more beams substantially improves the rejection ratio. As a rule of thumb, we find that the rejection ratio for 4 beams is approximately the square of the rejection ratio for 2 beams, for any rejection threshold. In all subsequent calculations we shall assume 4 beams are used.

Next we examine the declination dependence of the rejection ratio (fig. 6). We assume an observation frequency of 1420 MHz and consider two threshold levels, $N = 1$ dB and 3 dB. We also consider two hour angles, 0° (transit) and 45°. These results are easy to understand once you realize that the sidelobe velocity is proportional to the cosine of the declination and becomes stationary at the celestial pole (90° declination). Thus at high declinations the rejection ratio worsens while it is best near declination 0°. We find that at 1420 MHz and over a wide range of declination angles, a 3 dB RFI threshold will be very effective at discriminating RFI (rejection ratio ~ $10^{-5}$).

Figure 7 examines the frequency dependence of the rejection ratio. The sidelobe angular width varies inversely with the frequency. Thus at higher frequencies we average over a larger number of sidelobe levels, which improves the rejection ratio. We find that at the highest frequencies, even a 1 dB threshold level is sufficient to discriminate a large proportion of RFI.

We conclude this section by considering the ATA-32 configuration. Figure 8 shows the declination dependence of the rejection ratio at 1420 MHz. Because the spatial extent of ATA-32 is smaller than ATA-350, and because there are fewer antennas, the sidelobe pattern for ATA-32 is broader and less complex. Both of these factors reduce the effectiveness of pattern averaging. Even so, with a 3 dB rejection threshold and over most of the sky, the chances of false positive are typically 1 in 1000.

**Discussion**

    **Pattern Averaging**

We have seen that pattern averaging is very important for obtaining good rejection ratios. We emphasize that pattern averaging is effective only when the beam patterns are averaged *incoherently*. Suppose that the computing power were available to coherently FT over the entire 5 minute interval. In this case, the RFI still drifts through multiple sidelobes, but now sidelobe amplitudes are averaged instead of the powers. Calculations of this case show that the rejection ratio is no better than for a snapshot observation. The reason for this is that an average of many complex beam patterns is just another beam pattern. The antenna coefficients are a bit more complicated, but far from the synthetic maximum they have no particular relationship anyway. As a result, the far out sidelobe pattern is statistically the same whether the amplitudes are averaged or not.

The effectiveness of multibeam RFI rejection depends critically on the variance of the sidelobe distribution. If we consider a beamformer where the signal from each

element is summed with unit weight, the complex beam pattern $b$ at the point[1] $\vec{k}$ is given by

$$b(\vec{k}) = \sum_{i=1}^{N} e^{i\vec{k}\cdot\vec{r}_i} = \sum_{i=1}^{N} e^{i\phi_i}. \qquad (2)$$

Radio astronomical interferometers are frequently optimized to give a smooth distribution of baselines in the $(u, v)$ plane so that $\vec{r}_i$ and $\phi_i$ are essentially random variables (true for the ATA). Eq. (2) is therefore analogous to a 2-D random walk with $N$ unit-length steps. (It is also analogous to what is known as the Wilson distribution in x-ray crystallography.) From the central limit theorem, the sidelobe distribution converges in this case to a 2-D Gaussian distribution:

$$P(b) = \frac{1}{\pi N} \exp\left(-\frac{|b|^2}{N}\right). \qquad (3)$$

Thus it has a similar form on any random array provided that a) we are far from the main beam and b) we consider sidelobes sufficiently separated in $\vec{k}$ so that $\phi_i$ is a random variable. This bodes well for the application of our method on other radio astronomical interferometers.

The key to the success of pattern averaging is that it narrows the variance of the sidelobe distribution. To see how this happens, notice that the degree of narrowing depends on correlations between closely spaced points in the beam pattern. If we consider a track over a time period $\Delta t$, then $\vec{k}_{RFI}(t)$ varies as the target moves. From Eq. (2) we conclude that two sidelobe values $b(\vec{k}_{RFI}(t_1))$ and $b(\vec{k}_{RFI}(t_2))$ will be approximately independent if $\langle (\vec{k}_{RFI}(t_2) - \vec{k}_{RFI}(t_1)) \cdot r_i \rangle \geq 2\pi$. Since the variance $\sigma^2$ decreases as the inverse of the number of independent sidelobe values, we may write

$$\sigma^2 \propto \frac{2\pi}{\langle \Delta \vec{k}_{RFI} \cdot r_i \rangle}, \qquad (4)$$

where $\Delta \vec{k}_{RFI} = \vec{k}_{RFI}(t + \Delta t) - \vec{k}_{RFI}(t)$.

---

[1] The wavevector $\vec{k} \equiv k\hat{k}$, where $k = \dfrac{2\pi f}{c}$ and $\hat{k}$ is the direction. Also, $s = |b|^2$.

Because the sidelobes are intrinsically linked to the geometry of the array, this suggests an alternative optimization criterion for the array geometry: to minimize the variance of the sidelobes. In fact, ATA350 has been optimized for (*u*, *v*) coverage followed by an optimization to reduce the highest sidelobes in the vicinity of the synthetic maximum. The latter optimization tends to reduce the variance by a small amount. But an optimization scheme to reduce the global sidelobe variance might be worth considering, and it might prove beneficial for radio synthesis imaging as well.

**Comparison to Correlation Matrix Techniques**

One might ask how this method relates to other approaches for RFI identification. The most general analyses of the signal can be cast in terms of the correlation matrix of the signals arriving from all the interferometer elements (see Leshem 2000 for a review). For example, one can identify RFI by examining the largest eigenvalues of this matrix. If they are much larger than the signals expected from the target direction, then they must be RFI. In a more complicated extension of this technique, one could determine the direction of arrival of every eigenvector and compare this to the target direction, discarding signals not arriving from the target.

Here is another correlation matrix variant: Given a candidate signal, one can solve for the direction of arrival vector that maximizes the signal strength. If the direction of arrival does not agree with the target direction, then it cannot be arriving from the target. Our technique is an approximation to this method. Here we estimate the direction of arrival by generating an M-pixel "image" of the candidate signal and comparing it to the image of an idealized signal arriving from the target.

Correlation matrix techniques, however powerful, are distinct from the present method since they require a correlator as opposed to a few simple beamformers. Determining the direction of arrival also implies a time-consuming Fourier transform. Operationally, the present method is rather different from a correlation matrix approach.

**Limitations and Extensions**

The simulations here assume a perfect beamforming system. In practice, the complex gain factors applied to ATA beamformers will have some error, and other imperfections exist. Although these may have impact on the synthetic beam peak, they are not expected to change the statistics of the sidelobes. As discussed above, choosing any set of beamformer coefficients gives a similar distribution of sidelobe levels.

Here we have considered here only ground-stationary RFI whose radial vector from the array is perpendicular to the polar axis. The position of the RFI is another rich parameter space for the rejection ratio. Ground-based RFI in other directions or geostationary satellites will move through the sidelobe pattern more slowly than predicted here. Comparatively, LEO satellites and airplanes will move more quickly. Thus we expect lower rejection ratios for the former case and higher rejection ratios for the latter. Performing a serious theoretical evaluation of the average rejection ratio for all RFI would require knowledge of the distribution of RFI source type. This is beyond the scope of the present paper. The more modest goal of this paper is to give order of magnitude estimates and delineate the conditions under which multi-beam rejection might work. We conclude that as a technique it is promising.

### Conclusions

We have performed simulations to examine the efficacy of multi-beam discrimination of RFI in radio SETI. In models of the ATA we find that multi-beam discrimination is very effective at identifying RFI. Useful discrimination is obtained by looking for RFI in two simultaneous beams, but is substantially better if four beams are used. As might be expected, RFI discrimination is improved with increasing frequency or increasing array size. It is critically important that the RFI discrimination is made by incoherently averaging the signals from each beam over a period of time. Using 4 beams at 1420 MHz and a 5 minute integration period, the chance that RFI appears twice as large in one beam than in all the others is only 1 in $10^5$.

### Acknowledgements

We gratefully acknowledge Peter Backus and Peter Fridman for critical review of this manuscript. This research was supported by the SETI Institute.

### Appendix

Here we present a more detailed derivation of Equation (1) in the text. Suppose there are $M$ beams numbered 1, 2, …, $M$. The probability that beam 1 has a sidelobe level in the range $(s, s + ds)$ is

$$P(s)\, ds\,. \tag{A}$$

The probability that beam 2 has a sidelobe level in the range $(0, s/n)$ is

$$\int_0^{s/n} ds'\, P(s')\,. \tag{B}$$

The product of (A) and (B) is the probability that beam 1 has a level near *s* and beam 2 is at least *n* times lower than beam 1. Alternatively, we can phrase the second part as "beam 1 is at least *n* times larger than beam 2."

We also insist that beams 3, 4, …, *M* have low values and then integrate over all possible values for beam 1:

$$P_{1\,high,\ 2,3,...,M\,low} = \int_0^1 ds\, P(s) \left( \int_0^{s/n} ds'\, P(s') \right)^{M-1}. \qquad (D)$$

This is the probability that beam 1 has a level at least *n* times higher than all the others.

Up until now we have been quite specific about which beam has which value. While Eq. (D) is correct for the case we have considered, if beam 2 were higher than the others, we would still have a false alarm. Since there are M beams where the false alarm might appear, the probability that some unspecified beam is at least *n* times higher than all the others is

$$P_M(n) = M \int_0^1 ds\, P(s) \left\{ \int_0^{s/n} ds'\, P(s') \right\}^{M-1}. \qquad (1)$$

Finally, there is an interesting limiting case. If $n \to 1$, then $P_M(n=1)$ must be unity, because it is always true that either: a) there is a highest beam, or b) two or more beams tie for having the highest value. Returning to Fig. 4, we indeed find that $P_M(N=0) \to 1$ for all values of *M*.

Figures

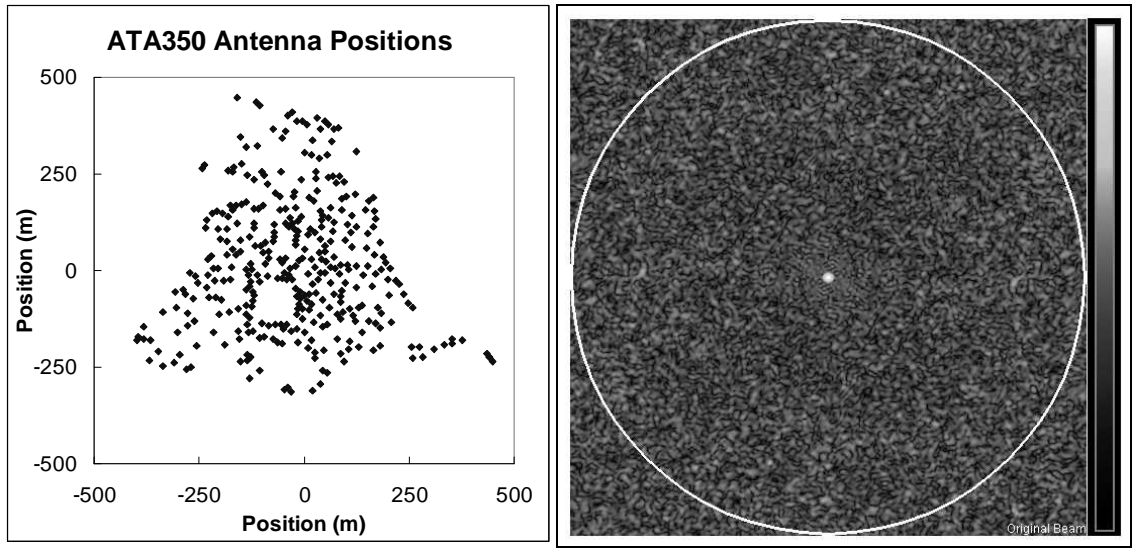

Figure 1: Left: Proposed layout of ATA-350 antennas. Right: Portion of the synthetic beam pattern of ATA-350 including the synthetic beam peak. The color bar identifies sidelobe power as measured from the maximum value in the image where white = 0 dB and black = -50 dB. The white circle shows the ATA primary beam FWHM.

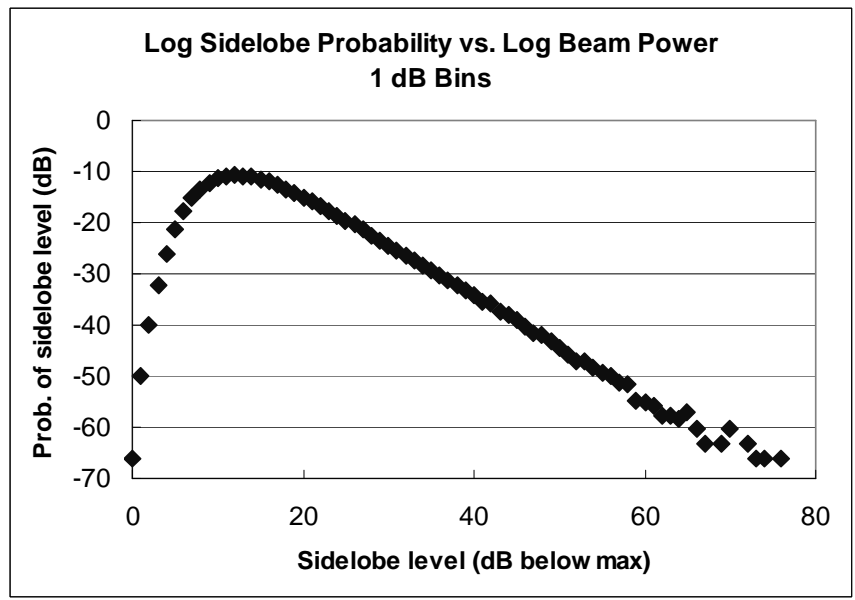

Figure 2: Numerical estimate of $P(s)$ in a snapshot observation. This is the probability that a randomly chosen sidelobe will have a level within 0.5 dB of a specified level, as a function of sidelobe level. Levels are measured from the maximum sidelobe level observed (excluding the beam maximum).

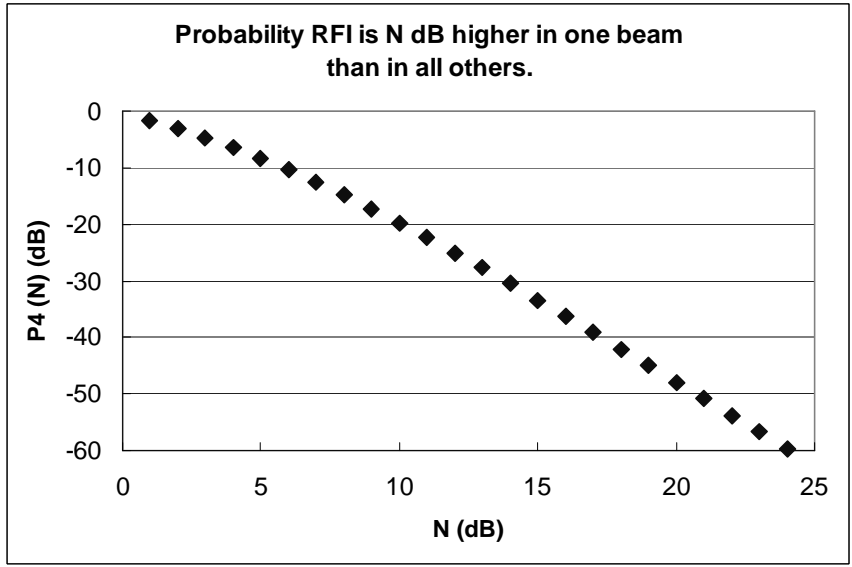

Figure 3: In a snapshot observation with four randomly placed observation beams, this graph shows the probability that one beam will exhibit RFI at least N dB higher than the same RFI in the other three beams. For example, the RFI will appear in one beam at least 10 dB higher than in all other beams about 1% of the time (P ~ -20 dB).

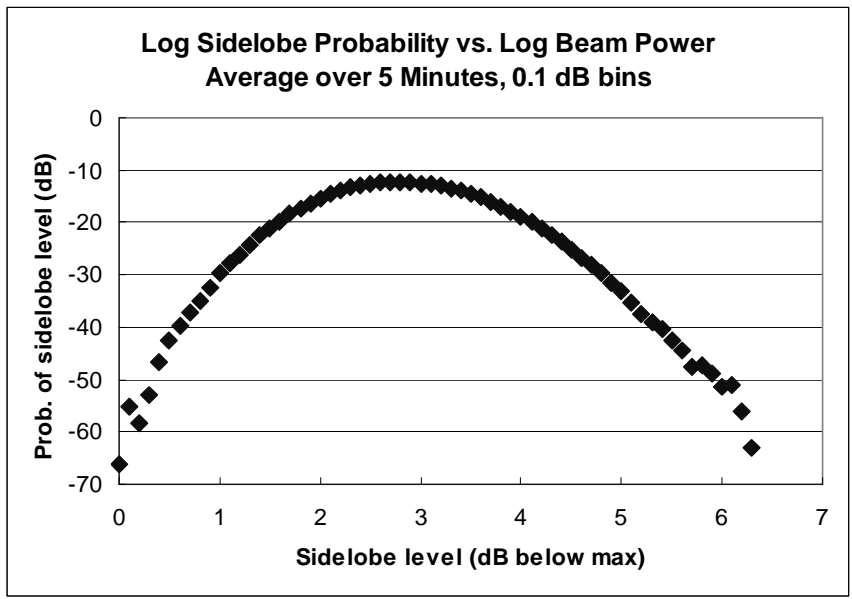

Figure 4: The probability that a randomly chosen sidelobe will have a level within 0.05 dB of a specified level, as a function of sidelobe level. This may be compared with fig. 2, but notice that the chosen bin width is different for the two figures. This calculation results when the beam power at the RFI position is averaged over a 5 minute time period (source declination 20°).

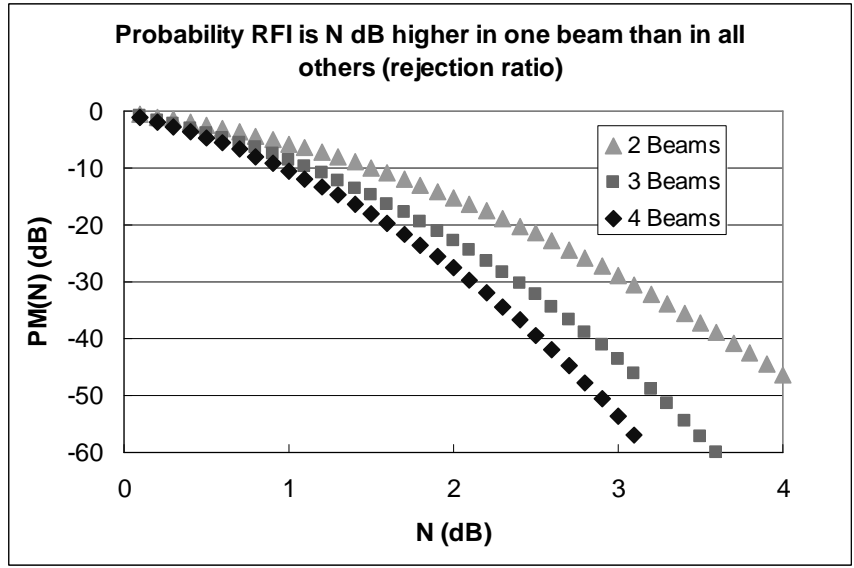

Figure 5: Similar to fig. 3, the probability that RFI will appear at least N dB higher in one beam than in all others, but for a 5 minute track. The blue curve is most comparable to fig. 3 and assumes 4 beams are used. The magenta and green curves show the same calculation for 2 and 3 beams, respectively.

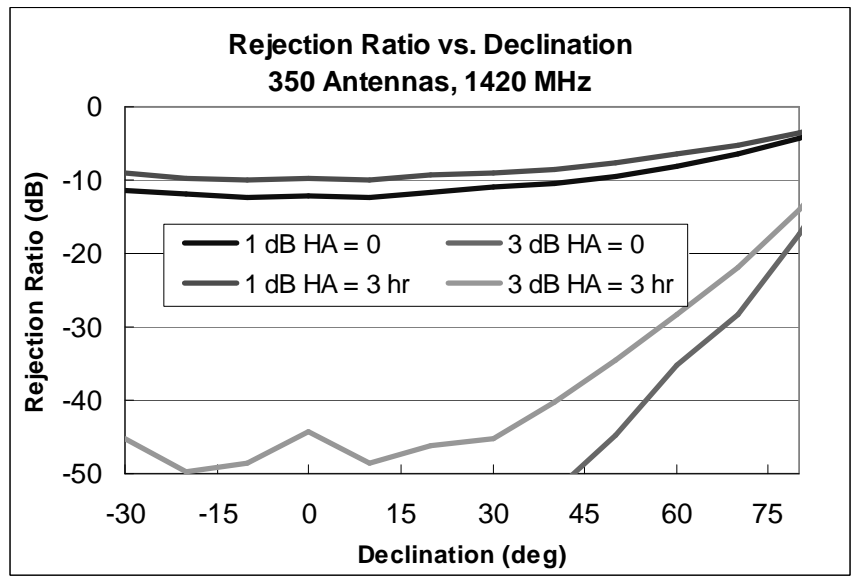

Figure 6: Rejection ratio as a function of declination. Ratios are shown for two discrimination thresholds (1 dB and 3 dB) and for two hour angles (0° = transit, and 45°). The rejection ratio gets noisy at very low levels due to the finite number of sidelobe samples in the calculation.

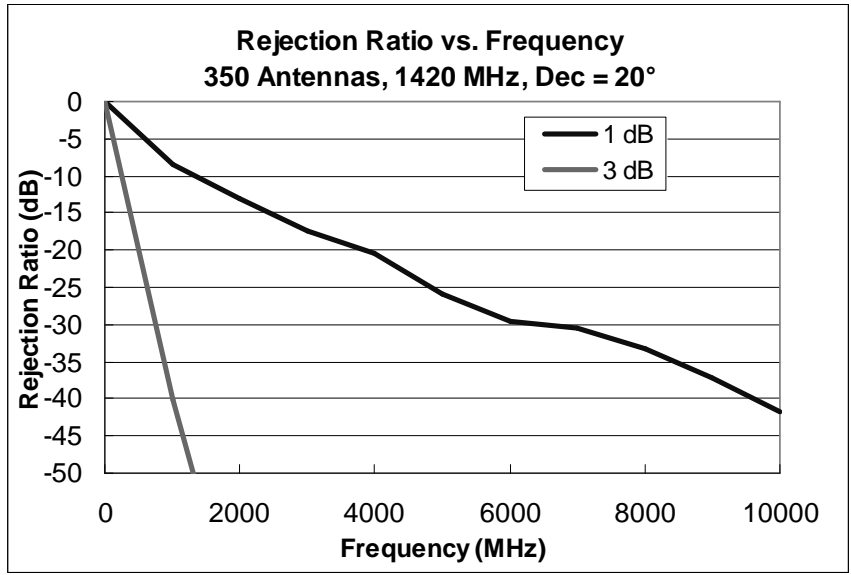

Figure 7: Rejection ratio as a function of frequency. As the frequency increases, the sidelobes become smaller and the rejection ratio improves.

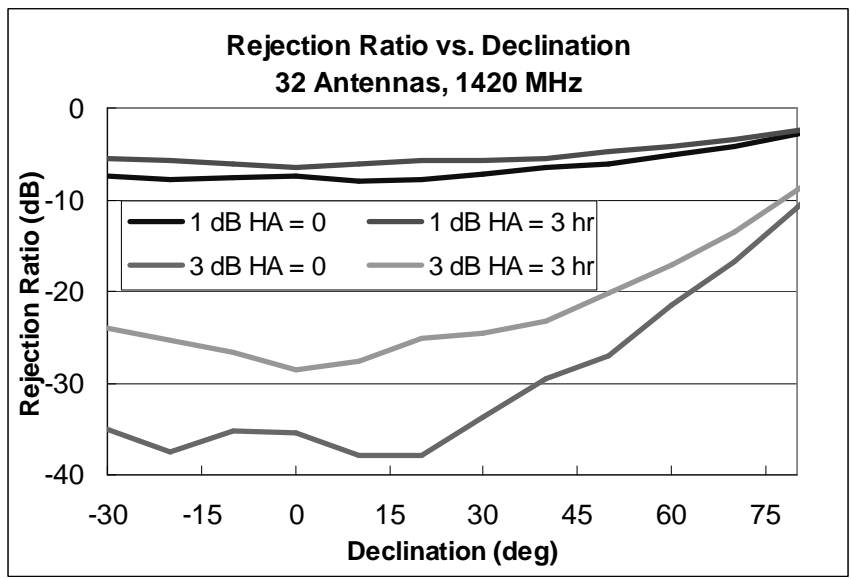

Figure 8: Rejection ratio as a function of declination for the ATA-32 configuration. Because there are fewer antennas and the spatial extents of the array are smaller than ATA-350, the rejection ratio is lower than for ATA-350, but still useful.